\documentclass{article}
\usepackage{spconf,amsmath,graphicx,hyperref}
\usepackage{algorithm}
\usepackage{algorithmic}
\usepackage{multirow}

\title{D4PM: A Dual-branch Driven Denoising Diffusion Probabilistic Model with Joint Posterior Diffusion Sampling for EEG Artifacts Removal}
\name{Feixue Shao\textsuperscript{1}, Xueyu Liu\textsuperscript{2}, Yongfei Wu\textsuperscript{2}, Jianbo Lu\textsuperscript{3}, Guiying Yan\textsuperscript{4}, Weihua Yang\textsuperscript{1}*\thanks{*Corresponding author.\\This work has been submitted to the IEEE for possible publication. Copyright may be transferred without notice, after which this version may no longer be accessible.}}
\address{\textsuperscript{1}School of Mathematics, Taiyuan University of Technology, China\\\textsuperscript{2}College of Artificial Intelligence, Taiyuan University of Technology, China\\\textsuperscript{3}National Human Genetics Resource Center, National Research Institute for Family Planning, China, \\\textsuperscript{4}Academy of Mathematics and Systems Science, Chinese Academy of Sciences, \\University of Chinese Academy of Sciences, China}
\begin{document}
\ninept
\maketitle
\begin{abstract}
Artifact removal is critical for accurate analysis and interpretation of Electroencephalogram (EEG) signals. Traditional methods perform poorly with strong artifact-EEG correlations or single-channel data. Recent advances in diffusion-based generative models have demonstrated strong potential for EEG denoising, notably improving fine-grained noise suppression and reducing over-smoothing. However, existing methods face two main limitations: lack of temporal modeling limits interpretability and the use of single-artifact training paradigms ignore inter-artifact differences. To address these issues, we propose D4PM, a dual-branch driven denoising diffusion probabilistic model that unifies multi-type artifact removal. We introduce a dual-branch conditional diffusion architecture to implicitly model the data distribution of clean EEG and artifacts. A joint posterior sampling strategy is further designed to collaboratively integrate complementary priors for high-fidelity EEG reconstruction. 
Extensive experiments on two public datasets show that D4PM delivers superior denoising. It achieves new state-of-the-art performance in EOG artifact removal, outperforming all publicly available baselines. \textit{The code is available at \href{https://github.com/flysnow1024/D4PM}{https://github.com/flysnow1024/D4PM}}. 
\end{abstract}
\section{Introduction}
\label{sec:intro}

EEG is a non-invasive technique for measuring brain activity with high temporal resolution, widely used in neuroscience and clinical diagnostics~\cite{cllmqw:22}, ~\cite{asj:23}. With the rapid advancement of mobile EEG devices, it is increasingly acquired in dynamic, uncontrolled environments, improving real-world applicability~\cite{lwzlscc:22}. However, EEG recordings are highly susceptible to contamination from non-cerebral physiological signals, particularly electrooculography (EOG), electromyography (EMG), and electrocardiography (ECG)~\cite{atmh:25}. Artifacts from eye movements, muscle activity, and cardiac rhythms overlap with neural signals in the time-frequency domain, compromising analysis and interpretation~\cite{cltl:19}. Therefore, robust and generalizable artifact removal methods are needed to preserve essential EEG components.

Extensive research has been investigated for EEG artifact removal, including regression models~\cite{gmvw:03}, blind source separation~\cite{zylz:21}, empirical mode decomposition (EMD)~\cite{zz:13}, and subspace reconstruction~\cite{yyqh:25}. To further improve artifact removal performance, a variety of hybrid approaches have emerged in recent years, such as wavelet-ICA~\cite{ilmm:07} and EMD-ICA~\cite{wtlz:15}. However, these traditional methods typically rely on multi-channel EEG recordings and manual expertise, making them difficult to generalize to diverse artifact scenarios~\cite{hlcc:25}. Moreover, their performance deteriorates significantly in single-channel settings~\cite{llz:19}. 

Compared to traditional methods, deep learning (DL) has gained significant attention in EEG denoising due to its powerful feature extraction and automatic representation learning capabilities. Various architectures, including CNNs~\cite{ssww:20}, FCNNs~\cite{ydf:18}, and Transformer-based models~\cite{pycmz:22}, have been explored, but they are prone to over-smoothing. While GAN-based methods mitigate this issue~\cite{yllqc:23}, they incur greater training complexity and unstable convergence. Recently, Huang et al~\cite{hllqc:24} applied diffusion models for EEG denoising, effectively addressing these limitations but sacrificing generalizability in more adverse degradation settings due to task-specific designs.

\begin{figure*}[htbp]
\centering
\includegraphics[width=0.9\textwidth]{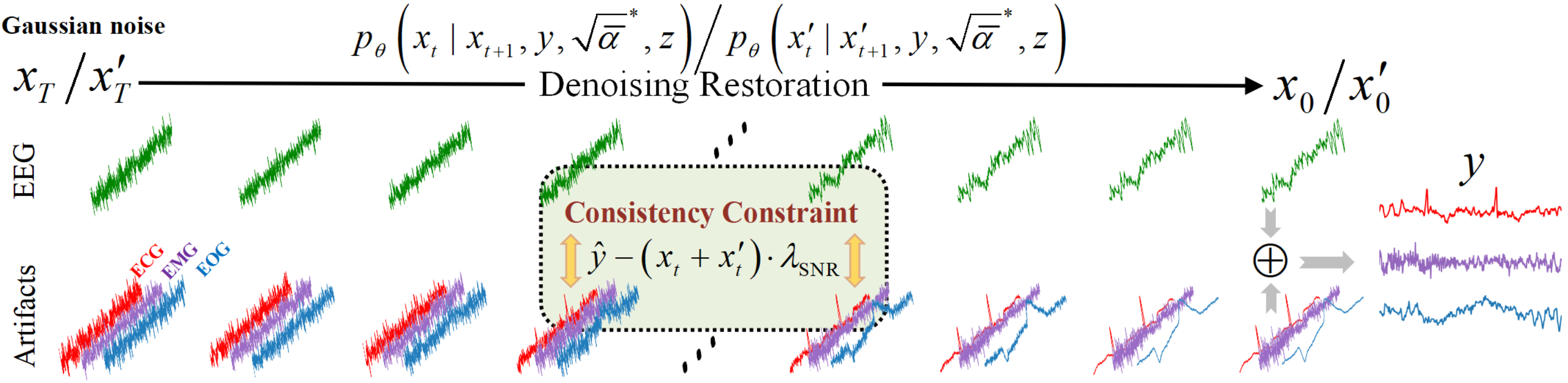}
\caption{Overview of the Joint Posterior Sampling in D4PM. The denoising restoration process $p(\cdot\,|\,\cdot)$ proceeds from left to right, where the top path represents the EEG branch $x_t$ and the bottom path denotes the artifact branch $x_t'$. Each denoising step is performed by its independently trained denoising network. At each step, data consistency is is promoted by the measurement model $y = x_t + x_t' \cdot \lambda_{\mathrm{SNR}}$.}
\label{fig1}
\end{figure*}
To address the aforementioned challenges, we propose a \textbf{D}ual-Branch \textbf{D}riven \textbf{D}enoising \textbf{D}iffusion \textbf{P}robabilistic \textbf{M}odel (\textbf{D4PM}) for multi-artifact removal in EEG.  
Unlike previous methods, our framework effectively captures both clean EEG representations and inter-artifact distinctions, demonstrating superior performance and robust generalization under diverse degradation conditions.  
To bridge the complementary nature of clean EEG reconstruction and artifact modeling, we formulate the artifact removal task as a joint posterior inference problem and design a collaborative sampling framework with consistency constraints, as shown in Figure~\ref{fig1}. The main contributions of this work are as follows:

\begin{itemize}
\item We formulate a unified generative framework for multi-type artifact removal on a synthetically constructed mixed dataset, aiming to enhance the model's capacity to capture the specificity among different artifact types.
\item The proposed D4PM framework enables two independently trained diffusion models to implicitly learn the underlying distributions of clean EEG and artifacts, capturing their respective temporal dynamics and structural patterns, and improving the interpretability of the model.
\item To leverage the complementary nature of the learned generative priors, we design a joint posterior sampling algorithm that imposes data consistency constraints during inference, enabling collaborative generation from the two conditional diffusion models.
\end{itemize}

\begin{figure}[t]
\centering
\includegraphics[width=\columnwidth]{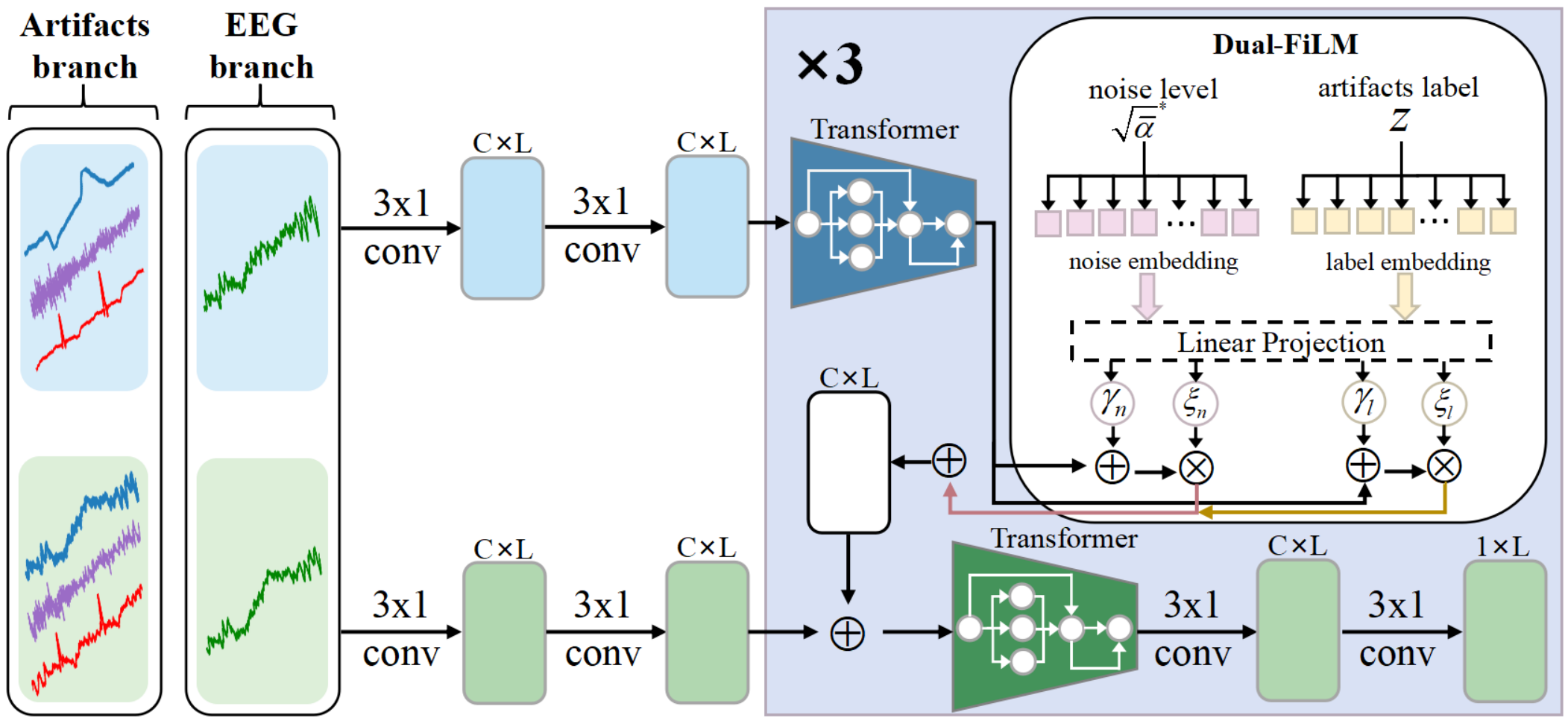}
\caption{The architecture of the denoising network. The two branches are trained in parallel with no weight sharing. Each branch consists of a dual-path model, where the two paths are modulated via a shared Dual-FiLM. The notation ``x3'' indicates that the corresponding colored module is repeated three times sequentially.}
\label{fig2}
\end{figure}
\section{METHODS}
\label{sec:format}
\subsection{Multi-conditional DDPM}
Conditional diffusion models have shown strong performance in various time series tasks.~\cite{tsse:21}, ~\cite{plk:21}. Therefore, we integrate additional conditioning information into the original diffusion process. Specifically, the diffusion model is trained to estimate the conditional distribution $p_\theta(x_0 \mid y)$, which approximates the true posterior $q(x_0 \mid y)$ for denoising purposes. Using discrete time step $t$ directly to condition the denoising network may lead to suboptimal performance and instability. To address this, we introduce a continuous noise-level variable $\sqrt{\bar{\alpha}_t} = \sqrt{\prod_{n=1}^t \alpha_n}$ as an embedded conditioning signal, allowing smoother guidance during training and inference. Specifically, ${\bar{\alpha}_t}^*$ is uniformly sampled from the interval $(\sqrt{\bar{\alpha}_{t-1}}, \sqrt{\bar{\alpha}_t})$. Furthermore, to capture artifact-specific characteristics, we use a class label $z$ as a categorical condition variable. 

Specifically, to better model EEG–artifact interactions, we simultaneously train two diffusion models on two paired data: one reconstructs clean EEG, the other learns artifact dynamics. For both branches, the training input is the noisy signal $y$. The EEG branch uses the pair $(x, y)$, where $x$ is the clean EEG, to train the denoising network $\epsilon_\theta$. The artifact branch uses $(x', y)$, where $x'$ denotes the artifact component, to train the artifact-aware denoising network $\epsilon'_\theta$. Let $x, x', y \in \mathbf{R}^N$, where $N$ representing the length of each EEG segment. In the EEG branch, the loss of our multi-conditional diffusion model is formulated as:
\begin{equation}
    \begin{split}
        \mathcal{L}_{\text{EEG}} &= \mathbf{E}_{x_0, y, \sqrt{\bar{\alpha}_t}^*, \epsilon, z} \\& \bigg[
     \left\| \epsilon - 
    \epsilon_\theta\left(
    \sqrt{\bar{\alpha}_t} \, x_0
    + \sqrt{1 - \bar{\alpha}_t} \, \epsilon, 
    y, \sqrt{\bar{\alpha}_t}^*, z \right) \right\|_1 \bigg].
    \end{split}
\label{eq:conditional_loss}
\end{equation}

For the artifact branch, replace input $x$ with $x'$. To enhance denoising, we introduce a dual-path denoising network with a shared dual feature-wise linear modulation (Dual-FiLM), as shown in Figure~\ref{fig2}. The EEG branch, as an example, processes the noise-perturbed signal $x_t$ and noisy EEG signal $y$ using identical architectures. Features are extracted via a shallow convolutional stack followed by three repeated Transformer encoders to model global dependencies. Dual-FiLM embeds noise level and artifact labels to generate channel-wise scaling ($\gamma$) and shifting ($\xi$) parameters for feature modulation. These are applied to enable class-aware multi-scale modeling and cross-path fusion. The fused features are projected through convolutional layers to output the denoised signal.

\subsection{Joint Posterior Sampling}

\begin{algorithm}[tb]
\caption{Joint Posterior Diffusion Sampling}
\label{alg:pattrain2}

\textbf{Input}: Noisy EEG $y$, trained EEG denoising network $\epsilon_{\theta}$, trained artifact denoising network $\epsilon_{\theta}'$, noise schedule $V = \{\beta_1, \ldots, \beta_T\}$, class label $z$, data consistency weight $\lambda_{dc}$ \\
\textbf{Output}: Denoised EEG $x_0$

\begin{algorithmic}[1]
\STATE Sample $y \sim p(y)$
\STATE Sample $x_T \sim \mathcal{N}(0, I)$
\STATE Compute $\alpha_t = 1 - \beta_t$
\STATE Compute $\bar{\alpha}_t = \prod_{n=1}^t \alpha_n$
\STATE Sample $\sqrt{\bar{\alpha}_t}^* \sim \text{Uniform}(\sqrt{\bar{\alpha}_{t-1}}, \sqrt{\bar{\alpha}_t})$

\FOR{$t = T, T-1, \ldots, 1$}
    \STATE Sample $\eta \sim \mathcal{N}(0, I)$ \textbf{if} $t > 1$, \textbf{else} $\eta = 0$
    
    \STATE Predict $x_0$ using EEG denoiser:
    \STATE \quad $x_0 = \frac{1}{\sqrt{\alpha_t}} \left( x_t - \frac{\sqrt{1 - \bar{\alpha}_t}}{\sqrt{\bar{\alpha}_t}} \cdot \epsilon_\theta(x_t, y, \sqrt{\bar{\alpha}_t}^*, z) \right)$

    \STATE Predict $x_0'$ using artifact denoiser:
    \STATE \quad $x_0' = \frac{1}{\sqrt{\alpha_t}} \left( x_t' - \frac{1 - \alpha_t}{\sqrt{1 - \bar{\alpha}_t}} \cdot \epsilon_\theta'(x_t', y, \sqrt{\bar{\alpha}_t}^*, z) \right)$

    \STATE \COMMENT{Calculate data consistency residual}
    \STATE $r = y - (x_0 + x_0' \cdot \lambda_{\text{SNR}} )$

    \STATE \COMMENT{Apply consistency constraint}
    \STATE $\hat{x}_0 = x_0 + \lambda_{dc} \cdot r$
    \STATE $\hat{x}_0' = x_0' + (1 - \lambda_{dc}) \cdot r$

    \STATE \COMMENT{Reverse diffusion update}
    \STATE $\mu_t = \frac{\beta_t \sqrt{\bar{\alpha}_{t-1}}}{1 - \bar{\alpha}_t} \cdot \hat{x}_0 + \frac{(1 - \bar{\alpha}_{t-1}) \sqrt{\alpha_t}}{1 - \bar{\alpha}_t} \cdot x_t$
    \STATE $\mu_t' = \frac{\beta_t \sqrt{\bar{\alpha}_{t-1}}}{1 - \bar{\alpha}_t} \cdot \hat{x}_0' + \frac{(1 - \bar{\alpha}_{t-1}) \sqrt{\alpha_t}}{1 - \bar{\alpha}_t} \cdot x_t'$

    \STATE $\hat{x}_{t-1} = \mu_t + \sigma_t \cdot \eta$
    \STATE $\hat{x}_{t-1}' = \mu_t' + \sigma_t' \cdot \eta$
\ENDFOR
\STATE \textbf{return} $x_0$
\end{algorithmic}
\end{algorithm}

The noisy EEG signal $y$ is modeled as a linear combination of the clean EEG signal $x$ and the artifact signal $x'$, as follows:
\begin{equation}
y = x + x' \cdot \lambda_{\text{SNR}},
\label{equ4}
\end{equation}
where $\lambda_{\text{SNR}}$ is a scaling factor used to control the SNR level. The SNR is defined as:
\begin{equation}
\text{SNR} = 10 \log \frac{\|x\|_2^2}{\|\lambda_{\text{SNR}} \cdot x'\|_2^2}.
\end{equation}
Following the setting in~\cite{hllqc:24}, the noisy EEG is generated with SNR ranging from $-5$ to $5$ dB by tuning the scaling factor $\lambda_{\text{SNR}}$, which controls artifact intensity. We formulate the artifact removal problem as a source separation task, where the objective is to recover the clean EEG from the observed noisy EEG. This inverse problem is typically ill-posed, as there exist multiple possible solutions that satisfy the formulation in Equation~\ref{equ4}. To simultaneously estimate the optimal $x$ and $x'$, we formulate the artifact removal task as a joint posterior sampling process inspired by Bayesian source separation. The joint posterior distribution is defined as:
\begin{equation}
p(x, x' \mid y) \propto p(y \mid x, x') \cdot p(x) \cdot p(x'),
\end{equation}
where $p(x)$ and $p(x')$ represent the prior distributions of the clean EEG and the corresponding artifact, respectively, while $p(y \mid x, x')$ denotes the observation likelihood, which measures the consistency between the noisy EEG and the estimated sources. During each iteration of the diffusion sampling process, we simultaneously perform the reverse steps of two independent diffusion models to predict $x_0$ and $x_0'$. Then compute the observation consistency residual $r$ and apply the consistency constraint on the predicted results $\hat{x}_{0}$ and $\hat{x}_{0}'$, as described in Algorithm~\ref{alg:pattrain2}. The $\lambda_{\text{dc}} \in [0,1]$ controls the directional weighting of the consistency residual. The key idea of this mechanism is to embed the generation of both the clean and artifact signals into a joint denoising framework under a shared noise model. By maintaining their respective priors, this enables collaborative optimization of both signal types. Finally, the corrected estimates $\hat{x}_0$ and $\hat{x}_0'$ are treated as the recovered clean EEG and corresponding artifact, respectively.

\begin{figure}[t]
\centering
\includegraphics[width=\columnwidth, height=0.25\textheight]{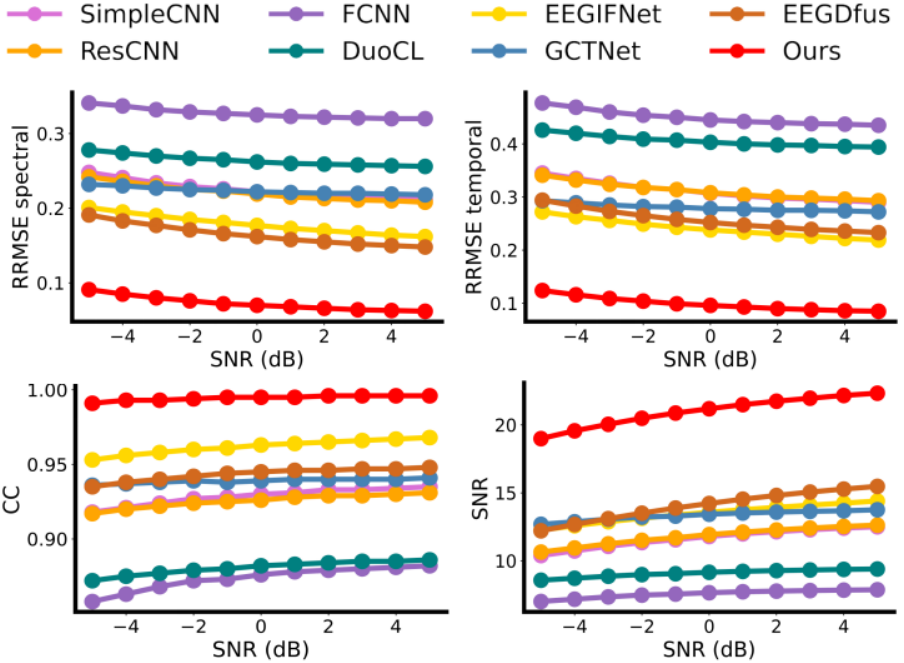}
\caption{Performance comparison of different baseline methods for EOG artifact removal under various SNR (dB) levels, evaluated by RRMSE\textsubscript{t}, RRMSE\textsubscript{s}, CC, and SNR.}
\label{fig3}
\end{figure}

\section{Experiment Results}
\label{sec:pagestyle}
\subsection{Datasets}
We conducted experiments on two different datasets: the EEGDenoiseNet and the MIT-BIH Arrhythmia dataset~\cite{gag:00}. To construct mixed dataset, we collected a total of 4514 EEG segments, 3400 EOG segments, 5598 EMG segments, and 3600 ECG segments for generating paired samples.
Specifically, for EMG data, following the strategy from previous studies, we randomly reused a portion of EEG segments during training to pair with EMG samples. The mixed dataset is divided into training (80\%), validation (10\%) and test (10\%) sets.

\subsection{Comparison with baseline methods}

We present a quantitative comparison between D4PM and several representative DL-based baselines, including FCNN~\cite{ydf:18}, SimpleCNN~\cite{ydf:18}, 1D-ResCNN~\cite{ssww:20}, NovelCNN~\cite{zwzl:21}, DuoCL\cite{gctm:22}, EEGIFNet\cite{pycmz:22}, GCTNet\cite{yllqc:23}, and EEGDfus\cite{hllqc:24}. To comprehensively evaluate the performance of all baseline methods, we conducted two types of training under varying SNR levels: individual models trained separately for each artifact type (as shown in Table~\ref{tab1}), and a unified model trained on a mixed dataset constructed in this work that includes all artifact types (as shown in Table~\ref{tab2}). The evaluation was conducted using five metrics: relative root mean square error in the time domain (RRMSE\textsubscript{t}), relative root mean square error in the spectral domain (RRMSE\textsubscript{s}), correlation coefficient (CC), SNR, and statistical significance measured by the \emph{p}-value.

\begin{table}[htbp]
\caption{Average RRMSE\textsubscript{t}, RRMSE\textsubscript{s}, CC, and SNR of different baseline methods on each artifact removal task. All CC values are statistically significant ($p < 0.05$). The best performance is highlighted in \textbf{bold}, and the second-best results are indicated by \underline{underline}.}
\centering
\scriptsize
\setlength{\tabcolsep}{0.25mm}
\begin{tabular}{llcccccccc}
\hline
Artifact & Metrics & FCNN & SimpleCNN & DuoCL & EEGIFNet & GCTNet & EEGDfus & \textbf{Ours}\\
\hline
\multirow{4}{*}{EOG} 
 & RRMSE\textsubscript{t} ↓ & 0.451 & 0.313  & 0.407 & \underline{0.241} & 0.281 & 0.257 & \textbf{0.098} \\
 & RRMSE\textsubscript{s} ↓  & 0.328 & 0.226 & 0.265 & 0.179 & 0.224 & \underline{0.165} & \textbf{0.072} \\
 & CC ↑   & 0.873 & 0.928  & 0.880 & \underline{0.962} & 0.939 & 0.943 & \textbf{0.995} \\
 & SNR ↑  & 7.551 & 11.605  & 9.069 & 13.48 & 13.315 & \underline{14.071} & \textbf{20.983} \\
\hline
\multirow{4}{*}{EMG} 
 & RRMSE\textsubscript{t} & 0.384 & 0.515  & 0.366 & 0.354 & \textbf{0.316} & 0.343 & \underline{0.321} \\
 & RRMSE\textsubscript{s}  & 0.269 & 0.332  & 0.262 & \underline{0.260} & \textbf{0.249} & 0.264 & \underline{0.260} \\
 & CC   & 0.904 & 0.862  & 0.904 & 0.926 & \underline{0.931} & 0.906 & \textbf{0.938} \\
 & SNR  & 9.185 & 6.211  & 10.047 & 9.612 & \textbf{11.736} & 11.307 & \underline{11.348} \\
\hline
\multirow{4}{*}{ECG} 
 & RRMSE\textsubscript{t}  & 0.522 & 0.353 & 0.419 & \underline{0.290} & 0.299 & 0.364 & \textbf{0.230} \\
 & RRMSE\textsubscript{s} & 0.372 & 0.257  & 0.277 & \underline{0.209} & 0.240 & 0.295 & \textbf{0.169} \\
 & CC   & 0.837 & 0.923  & 0.889 & \underline{0.948} & 0.945 & 0.864 & \textbf{0.966} \\
 & SNR   & 6.082 & 9.958  & 8.362 & 11.609 & \underline{12.086} & 11.206 & \textbf{14.128} \\
\hline
\end{tabular}
\label{tab1}
\end{table}
As shown in Table~\ref{tab1}, for EOG and ECG artifact removal tasks, D4PM outperforms all existing baseline methods across all evaluation metrics. Even though our model is trained under a mixed dataset setting, while baselines are trained specifically on single-artifact datasets.
Notably, in the EOG removal task under the challenging SNR level of $-5$ dB, D4PM achieved a CC of 0.991, a RRMSE\textsubscript{t} of 0.124, a RRMSE\textsubscript{s} of 0.091, and an SNR of 18.994, significantly surpassing the performance of other models even under the 5 dB condition, as illustrated in Figure~\ref{fig3}. These results demonstrate the remarkable robustness of D4PM against severe EOG artifacts.

For mixed dataset, Table~\ref{tab2} summarizes the average performance across the three artifact types. The results demonstrate that D4PM achieves the most competitive overall performance.

\subsection{Ablation studies}

To evaluate the effectiveness of each key component in the proposed D4PM model, we conducted ablation experiments on the mixed dataset with the following three configurations:
\begin{itemize}
    \item \textbf{Conditional DDPM (Base):} A conditional diffusion model with noise scheduling and noisy EEG, where only the EEG branch is used for prediction.
    \item \textbf{Base + Artifacts:} Builds upon the base model by introducing an artifact branch and performing joint posterior sampling based on both EEG and artifact features.
    \item \textbf{D4PM (Our Method):} Extend the previous configuration by incorporating artifacts label embeddings.
\end{itemize}
As shown in Table~\ref{tab3}, compared to Conditional DDPM, adding the artifact modeling branch significantly improves the CC values across all artifact types, with especially pronounced gains in the EMG task. This demonstrates the importance of explicitly modeling artifact information when dealing with complex contamination. D4PM achieves superior or competitive performance across most evaluation metrics in all three artifact types, and consistently outperforms other variants on every evaluation metric. These results confirm the effectiveness of the proposed design.
We attribute the performance gains to two complementary factors: the integration of noise-level and artifacts-label embeddings, which facilitates the modeling of inter-artifact distributional variability, and the proposed joint posterior sampling mechanism, which enables effective fusion of EEG and artifact-specific representations.

\begin{table}[htbp]
\caption{Average results of each metric on the mixed dataset compared with other baseline methods. All CC values are statistically significant ($p < 0.05$). The best results are highlighted in \textbf{bold}, and the second-best results are \underline{underlined}.}
\centering
\footnotesize
\setlength{\tabcolsep}{2mm}
\begin{tabular}{l|cccc}
\hline
Method & RRMSE\textsubscript{t} ↓  & RRMSE\textsubscript{s} ↓  & CC ↑ & SNR ↑  \\
\hline
1D-ResCNN     & 0.416 & 0.285 & 0.899 & 8.471 \\
SimpleCNN     & 0.404 & 0.281 & 0.903 & 8.757 \\
FCNN         & 0.356 & 0.251 & 0.911 & 10.123 \\
DuoCL        & 0.312 & 0.209 & 0.931 & 11.331 \\
EEGIFNet     & 0.282 & 0.210 & 0.950 & 12.028 \\
NovelCNN     & 0.260 & 0.203 & 0.946 & 13.687 \\
GCTNet      & \underline{0.226} & \underline{0.171} & \underline{0.956} & 15.100 \\
EEGDfus     & 0.239 & 0.195 & 0.920 & \textbf{16.114} \\
\textbf{Ours}             & \textbf{0.217} & \textbf{0.168} & \textbf{0.966} & \underline{15.487} \\
\hline
\end{tabular}
\label{tab2}
\end{table}

\begin{table}[h]
\caption{Ablation study on different components, and the best results are shown in \textbf{bold}.}
\centering
\footnotesize
\setlength{\tabcolsep}{1.5mm}
\begin{tabular}{@{}clccc@{}}
\hline
\multirow{3}{*}{Artifact} & \multirow{3}{*}{Metric} & \multicolumn{3}{c}{Method} \\
\cline{3-5}
 & & Conditional & Base+ & Our \\
 & & DDPM(Base) & Artifacts & Method \\
\hline
\multirow{4}{*}{EOG} 
 & RRMSE\textsubscript{t} ↓  & \textbf{0.099} & 0.105 & \textbf{0.099} \\
 & RRMSE\textsubscript{s} ↓  & \textbf{0.071} & 0.077 & 0.073 \\
 & CC ↑      & 0.993 & 0.994 & \textbf{0.995} \\
 & SNR ↑     & \textbf{21.850} & 20.530 & 20.987 \\
\hline
\multirow{4}{*}{EMG} 
 & RRMSE\textsubscript{t} & 0.369 & 0.388 & \textbf{0.321} \\
 & RRMSE\textsubscript{s} & 0.313 & 0.332 & \textbf{0.260} \\
 & CC      & 0.848 & 0.904 & \textbf{0.938} \\
 & SNR   & 10.922 & 10.491 & \textbf{11.348} \\
\hline
\multirow{4}{*}{ECG} 
 & RRMSE\textsubscript{t} & 0.250 & 0.268 & \textbf{0.230} \\
 & RRMSE\textsubscript{s}  & 0.202 & 0.219 & \textbf{0.170} \\
 & CC     & 0.918 & 0.946 & \textbf{0.966} \\
 & SNR     & \textbf{15.570} & 13.807 & 14.128 \\
\hline
\multirow{4}{*}{Avg} 
 & RRMSE\textsubscript{t} & 0.239 & 0.254 & \textbf{0.217} \\
 & RRMSE\textsubscript{s}  & 0.195 & 0.209 & \textbf{0.168} \\
 & CC      & 0.920 & 0.948 & \textbf{0.966} \\
 & SNR     & \textbf{16.114} & 14.942 & 15.487 \\
\hline
\end{tabular}
\label{tab3}
\end{table}

\begin{figure}[t]
\centering
\includegraphics[width=\columnwidth, height=0.25\textheight]{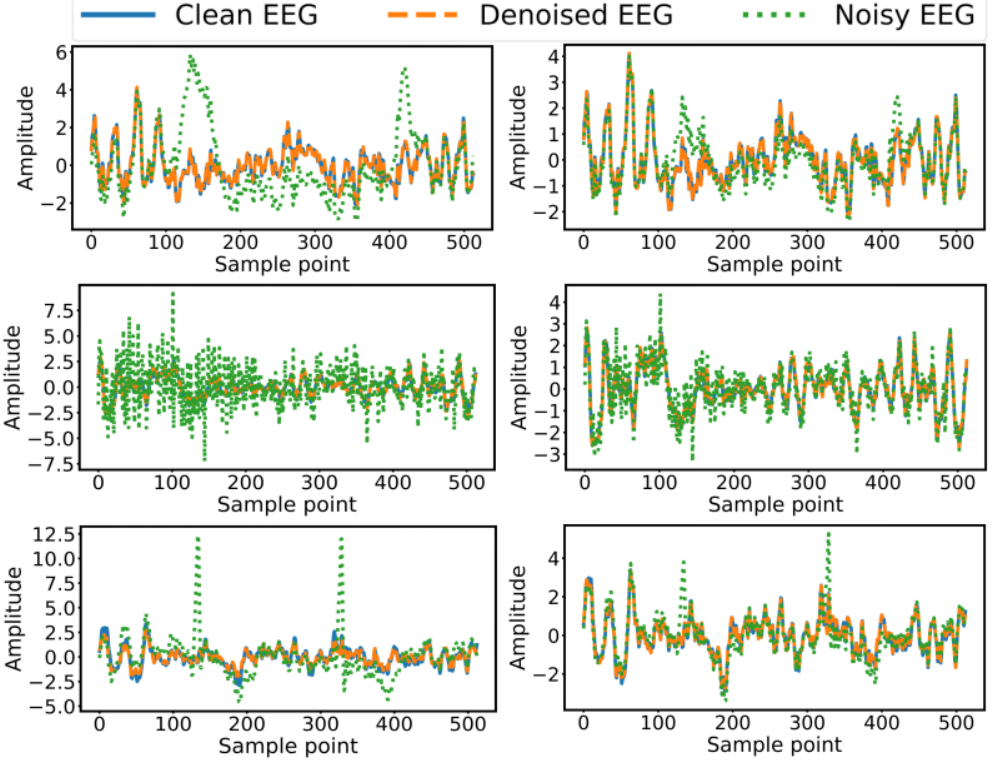}
\caption{Comparison of clean, noisy, and denoised EEG signals across artifact types and SNR levels(rows: EOG, EMG, ECG; columns: SNR = $-5$ dB and SNR = $5$ dB).}
\label{fig4}
\end{figure}
\subsection{Visualization}
To intuitively validate the denoising capability of the proposed method, noisy EEG signal samples were selected from mixed dataset, and their waveforms before and after denoising were compared, as illustrated in Figure~\ref{fig4}. Despite the presence of large-amplitude noise in certain samples, the proposed D4PM demonstrates a strong capability in suppressing noise interference while preserving essential neural activity components, thereby reconstructing high-fidelity EEG signals. Moreover, even in cases where the noise amplitude is minimal and the noisy EEG signals are visually similar to clean EEG, D4PM remains capable of maintaining signal integrity and retaining activity-related components without introducing oversmoothing or information loss.

\section{CONCLUSION}
\label{sec:foot}

In this work, we propose D4PM, a unified generative framework for multi-type artifact removal in EEG signals. We design a dual-branch architecture to disentangle and reconstruct EEG and artifact features separately, enhancing interpretability while improving fine-grained denoising performance. By leveraging data-driven generative modeling, D4PM demonstrates robust artifact removal on single-channel EEG, even under conditions of strong or weak EEG-artifact correlation, achieving faithful signal reconstruction. Generalization experiments on multi-channel datasets show that the proposed model consistently outperforms existing baselines on unseen and cross-domain samples. Future work will explore mixed-artifact modeling to further advance the practical deployment of this technology.



\bibliographystyle{IEEEbib}
\begin{small}

\bibliography{strings,refs}
\end{small}
\end{document}